A Physical source of dark energy and dark matter.

I. Gontijo

Jet Propulsion Laboratory, California Institute of Technology

4800 Oak Grove Drive, Pasadena, California, 91109-8099

igontijo@jpl.nasa.gov

ABSTRACT

A physical mechanism that produces three energy components is proposed as the common origin of dark energy and dark matter. The first two have equations of state W ~ 0 and act like dark matter, while the last has W ~ -1 at low redshifts making it a candidate for dark energy. These are used to model the supernovae Union2 data resulting in a curve fitting identical to the ΛCDM model. This model opens new avenues for Cosmology research and implies a re-interpretation of the dark components as a scalar field stored in the metric of spacetime.

PACS codes: 95.35.+d, 95.36.+x

                                         

I. INTRODUCTION

For over a decade mounting experimental evidence has given credence to the Concordance Cosmology or ΛCDM (Λ and Cold Dark Matter) model to describe the contents and evolution of the universe. After calibration of Type Ia Supernovae (SNIa) observations at increasing redshifts two groups [1] and [2] concluded that there was enough evidence for an accelerated expansion of the universe at low redshifts (since $z \sim 1$). This led to the postulation of a new and as yet undetected universal fluid now called dark energy that exerts negative pressure and causes acceleration according to General Relativity.[3]

A rich set of ideas on the nature of dark energy and its evolution (or not) have been proposed [4,5,6,7,8,9] and reviewed in recent books [10,11,12]. The two simplest interpretations of dark energy are the cosmological constant Λ introduced by Einstein or an exotic perfect fluid whose density might vary with time and redshift. It is almost a default expectation that the increasing quantity and precision of observational results would converge to one and only one of the dark energy models. This has happened to some extent but surprisingly the best model that fits the observations is still the simplest one – the cosmological constant Λ – effectively a new constant of nature, undiluted by the universal expansion. Other observational evidence such as galaxy clustering [13], Baryon Acoustic Oscillations [14], tomographic weak gravitational lensing [15] and the Cosmic Microwave Background (CMB) Anisotropies [16] also support the ΛCDM model and the latest Wilkinson Microwave Anisotropy Probe (WMAP7) seven years results

                                                                 2

[17] lead to a universe composed of ~4.6% baryonic matter, ~22.9% Cold Dark Matter and 72.5% Dark Energy.

It is well known that there are at least four major problems with the cosmological constant interpretation [11] namely a) - the "fine tuning" problem:120 orders of magnitude discrepancy with vacuum energy estimates - arguably the largest disagreement in all science between a theory and the previously established body of knowledge, b) - the coincidence problem: matter and dark energy densities coincide within a factor of 2 or 3, but no relation between them has been found, c) - conflict with inflation, which requires an evolving dark energy and d) - the lack of a physical mechanism despite over 10 years of intense investigations. These difficulties have been debated vigorously and led to questioning whether Dark Energy really exists or is an illusion.[18] Invoking the principle of ontological economy commonly attributed to William of Occam (*pluralitas non est ponenda sine necessitate*), one ought to enquire whether it is possible to find physical sources of dark energy and dark matter and avoid the much harder problems of finding new forms of matter/energy or new constants of nature to explain recent observational results. Taking as the starting point the General Relativity (GR) equations with a cosmological constant $\Lambda$:

$$R_{\mu\nu} - \frac{1}{2}g_{\mu\nu}R + \Lambda g_{\mu\nu} = 8\pi G T_{\mu\nu} \qquad (01)$$

The standard practice following Zeldovich [19] is to assume vacuum conditions ($T_{\mu\nu} = 0$) and move the $\Lambda$ term to the right hand side of eq. (01), so that it can be interpreted as the stress-energy tensor of vacuum: $T_{\mu\nu}^{vac} = -\Lambda g_{\mu\nu}$. If this approach is acceptable then a flow in the opposite direction, from the matter to the geometry side, should be allowed

                                                                                              3

too, provided that both sides of the equation are still independently covariant. If such a flow existed, it would manifest itself as energy loss from the $T_{\mu\nu}$ tensor with a corresponding change in the metric side of equation (01). It is interesting to note that the concept of a 'missing' energy density that still takes part in physical processes is of course well known in Physics: holes - missing electrons in the valence band of semiconductors - are quasiparticles that have mass, carry positive charge and form hydrogen-like atoms (excitons) when they bind with an electron. Holes are essential to explain the anomalous sign of the Hall coefficient in some metals and the transport properties in semiconductor devices.[20] In what follows we argue for a flow in the opposite direction to that proposed by Zeldovich, i.e., from the matter to the geometry side of the GR equations, thereby introducing a physical source of dark matter and dark energy. This will prove to be an intriguing approach, as it produces a fit to the SNIa Union2 data [21] that is as good as the ΛCDM model, is in general agreement with other observations and potentially can solve all four problems of the ΛCDM model.

## II. ENERGY DENSITIES & PHOTON REDSHIFTING

Particles and radiation in thermal equilibrium were present in the plasma era after inflation, in an extremely homogeneous and isotropic configuration. As the universe expanded and cooled, radiation decoupled from the particles and the continuing expansion caused it to redshift to the present day cosmic microwave background (CMB). This picture of the universe is well described by the Friedmann Lemaître Robertson Walker (FLRW) metric that simplifies the GR tensor equations to the two Friedmann's equations:

                                                                      4

$$H^2 = \left(\frac{\dot{a}}{a}\right)^2 = \frac{8\pi G}{3}\sum_i \rho_i - \frac{k}{a^2} \qquad (02)$$

and

$$\frac{\ddot{a}}{a} = -\frac{4\pi G}{3}\sum_i(\rho_i + 3p_i) \qquad (03)$$

Where H is the Hubble parameter, the scale factor is a(t), G is the gravitational constant, k = (0, +1 or -1) describes flat, closed and open geometries respectively and each matter species of density $\rho_i$ and pressure $p_i$ is treated as a perfect fluid. The ΛCDM model includes baryonic matter of density $\rho_{Baryon}$, cold dark matter $\rho_{CDM}$, CMB radiation and other relativistic particles $\rho_{rel}$ and dark energy $\rho_{DE}$ or the cosmological constant Λ. Equation (03) implies accelerated expansion for $p < -\rho/3$.

The CMB radiation has the lowest density at present, being redshifted by a factor of z~1090 since last scattering.[17] We begin by calculating how much energy density has been lost by the CMB radiation through the photon redshifting process. The CMB photon energy can be written as a function of the scale factor 'a' as:

$$E(a) = hc/\lambda = (hc/\lambda_{CMB}).(a_0/a) \qquad (04)$$

Where $a_0 = a(t=0)$. The energy lost per CMB photon due to expansion from the recombination era '$a_r$' to 'a' is

$$\Delta E(a) = E(a_r) - E(a) = (hc/\lambda_{CMB}).a_0.(1/a_r - 1/a) \qquad (05)$$

Now the photon density needs to be parameterized by 'a' as well. A black body spectrum of temperature T = 2.728K fits the measured CMB spectrum extremely well, allowing the calculation of the density of radiation modes resulting in n = $4.128 \times 10^8$ photons.m$^{-3}$. The

                                                                                                                      5

total number of photons in a bubble of present radius 'a$_0$' is therefore N = n.(4π/3).a$_0^3$. Assuming the total number of photons N is conserved, the photon density ρ$_r$(a) in an expanding bubble of volume V(a) = (4π/3).a$^3$ for any scale factor 'a' (a$_r$ ≤ a ≤ a$_0$) is ρ$_r$(a) = N/V(a) = n.(a$_0$/a)$^3$ photons/m$^3$. Finally, the CMB energy density lost due to expansion from a$_r$ to a is:

$$\rho_L(a) = \Delta E(a).\rho_r(a) = (n.hc/\lambda_{CMB}).(a_0^4/a^3).(1/a_r - 1/a) \tag{06}$$

and the dimensionless energy density Ω$_L$(a) is obtained by normalizing equation (06) to the critical density ρ$_c$ = (3c$^2$/8πG).H$_0^2$, so that

$$\Omega_L(a) = \Omega_r^o. (a_0/a)^4.(a/a_r - 1) \tag{07.a}$$

Or, converting from the scale factor 'a' to redshift according to (z+1) = a$_0$/a:

$$\Omega_L(z) = \Omega_r^o.(z + 1)^3.(z_r - z) \tag{07.b}$$

Where Ω$_r^o$ = [n.hc/(ρ$_c$.λ$_{CMB}$)] = 4.9x10$^{-5}$ is the present era CMB dimensionless energy density and (z$_r$ +1) = a$_0$/a$_r$. Eqs.(7) are plotted in Fig. 1 and it is clear that this energy density peaked shortly after the recombination era at a$_{max}$ = (4/3).a$_r$ and Ω$_L$(a$_{max}$) = (27/256). Ω$_r^o$. (a$_0$/ a$_r$)$^4$. Notice also that Ω$_L$(a$_0$) = Ω$_r^o$.(a$_0$/a$_r$ - 1) ~ 0.053 = 5.3%. This is remarkable, as it shows that the lost CMB energy density at present is larger than all the baryonic matter density in the ΛCDM model!

If Ω$_L$(a) is treated as a perfect fluid in cosmology then it is important to calculate its pressure and equation of state. The pressure is obtained by requiring Ω$_L$(a) to satisfy the local energy conservation law in GR:[3]

Copyright 2012. All rights reserved.                                                               6

$$\dot{\rho} = -3(\dot{a}/a).(\rho + p) \tag{08}$$

Equations (7.a) and (8) result in

$$p(a) = -(1/3).\Omega_r^o.\rho_c.(a_0/a)^4 \tag{09}$$

$p(a)$ is negative for all values of a in the range ($a_r \leq a \leq a_0$) and the equation of state for

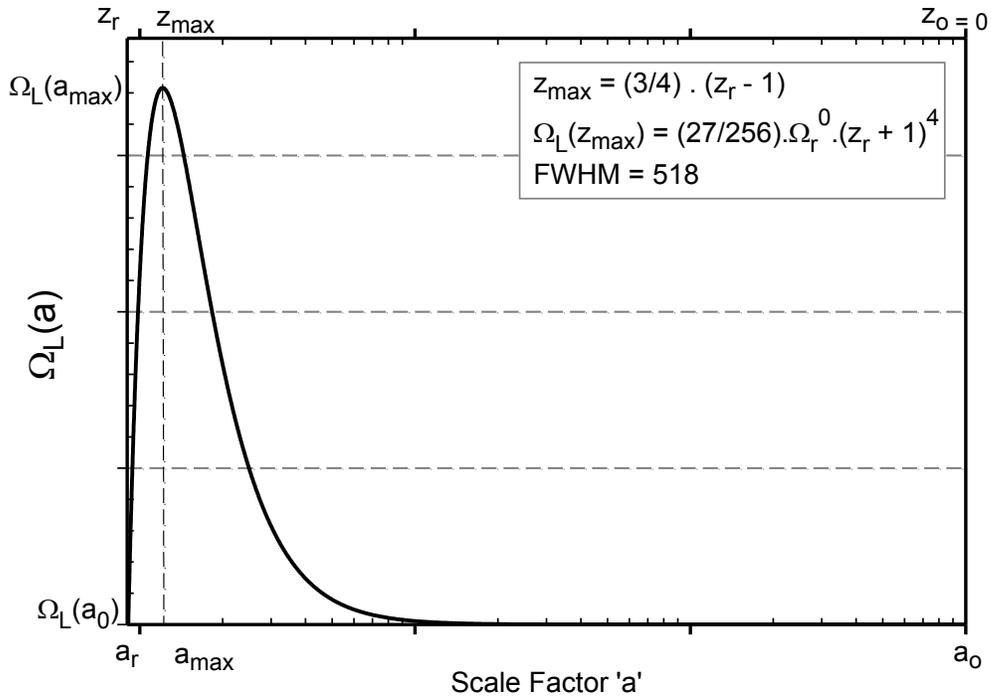

*Fig. 1 – Lost energy density as a function of the scale factor 'a', from the recombination era $a_r$ to the present era $a_0$. The redshift axis is obtained by making $(z+1) = a_0/a$.*

the CMB lost energy density is:

$$W_L(a) = p(a)/\rho(a) = -(1/3).a_r/(a-a_r), (a_r \leq a \leq a_0) \tag{10.a}$$

or

$$W_L(z) = - (1/3).(z+1)/(z_r-z), \quad (0 < z < z_r) \tag{10.b}$$

At present $W_L(a_0) = W_L(z=0) \sim -3 \times 10^{-4}$ which is virtually identical to zero, while $W_L(a_{max}) = W_L[(4/3).a_r)] = -1$, where $a_{max}$ is the scale factor corresponding to the peak of $\Omega_L(a)$ in Fig. 1. In terms of redshift, $a_{max}$ corresponds to $z_{max} = (3z_r-1)/4 = 817.25$ and (10.b) also leads to $W_L(z_{max}) = -1$ as expected. Also of interest is the value of $z_{avg}$, for which the equation of state $W_L(z)$, averaged over the redshift range from $z = 0$ to $z_{avg}$, produces a value of -1:

$$-1 = \frac{1}{z_{avg}} \cdot \int_0^{z_{avg}} w_L(z).dz \tag{11}$$

resulting in $z_{avg} = 1068.3$. Another important redshift is $z_{accel}$, defined such that $W_L(z_{accel}) = -1/3$. It can be obtained from eq. (10.b) and $z_{accel} = (z_r-1)/2 = 544.5$. If the energy density $\Omega_L(z)$ were allowed to gravitate on its own, it would produce accelerated expansion in the redshift range $z_{accel} < z < z_r$. In addition, assuming a flat universe (k=0), eq. (02) can be solved for the energy density $\Omega_L(a)$ either on its own or together with the radiation density $\Omega_r(a) = \Omega_r^o . (a_0/a)^4$. Either way the resulting time evolution of the scale factor is $a(t) = Q\, t^{2/3}$, where Q is a constant. This is the same time evolution produced by matter as discussed later.

So far we have seen that the energy density $\Omega_L(z)$ lost by the CMB photons since recombination would produce negative pressure for all values of $z < z_r$ and accelerated expansion (if it is dominant) in the range $z_{accel} < z < z_r$. According to the $\Lambda$CDM model the universe is 13.7 Billion years old and recombination happened very early, at around 380,000 years after the big bang. Our knowledge of the energy densities present earlier

 

during inflation is vague at best, so it is not unreasonable to assume that the dominant energy density during inflation consisted of extremely short wavelength photons. In this case the redshifting process would have transferred a large energy density from the photons into a density analogous to $\Omega_L(z)$ which could still be present today as a tail from inflation, whereas the inflation photons would have been absorbed during the plasma era. This density would have a peak similar to Fig. 1, but on a much shorter time scale. Therefore we define the energy density lost by the big bang photons, from inflation to the present day, in analogy to eqs. (7):

$$\Omega_I(a) = \Omega_I^o \cdot (a_0/a)^4 \cdot (a/a_I - 1) \tag{12.a}$$

$$\text{Or, } \Omega_I(z) = \Omega_I^o \cdot (z + 1)^3 \cdot (z_I - z) \tag{12.b}$$

Here the redshift and scale factor at inflation are $z_I$ and $a_I$ respectively and the present day energy density is $\Omega_I^o$. This function has interesting properties for the inflation period: given an inflation redshift $z_I$, $\Omega_I(z)$ reaches its peak very quickly at $(3z_I-1)/4$ and decays for lower values of z, suggesting a mechanism that could explain how inflation is turned OFF. As it has the same functional form of $\Omega_L(z)$, it produces an expansion that scales as $a(t) \propto t^{2/3}$, again identical to the baryonic matter behavior, with accelerated expansion for $z > (z_I-1)/2$.

Both energy densities $\Omega_L(z)$ and $\Omega_I(z)$ have the desirable properties of negative pressure and negative equations of state and would produce accelerated expansion in the absence of other densities, for part of their dilution histories. However, it is clear that on their own they cannot explain the SNIa data and the present cosmic accelerated expansion;

                                                                                                                                     9

these are events close to redshift z = 0, while those properties of $\Omega_L(z)$ and $\Omega_I(z)$ appear very early on (for redshifts larger than 544.5 for $\Omega_L(z)$ and vastly larger for $\Omega_I(z)$ ). At low redshifts both equations of state - although negative - tend to zero. In order to explain the present day accelerated expansion and to agree with other observational results, a different kind of energy density is needed, which is constant or slowly varying for low redshifts (say below z ~ 2) and with a negative equation of state close to w(z) = -1. It can be argued that an energy density with these properties has already been detected experimentally as a consequence of the cosmic UV luminosity density and star formation rate since the re-ionization epoch.[22,23] Unlike the inflation and recombination events which were treated as happening in a very short time scale so that they could be modeled as taking place at a single redshift ($z_i$ and $z_r$ respectively), the star formation rate and consequent photon production are events that started at around a re-ionization redshift of $z_{reI}$ ~ 20, peaked at $z_{peak}$ ~ 2.4 and are still in progress, producing the Extra-galactic Background Light.[24] The energy density $\Omega'_{reI}(z)$ lost by these continuously produced and redshifted photons is given by:

$$\Omega'_{reI}(z) = \Omega'_{reI,0} \int_z^{z_{reI}} SFR(z) \cdot \frac{z}{(1+z)^2 H(z)} \, dz \qquad (13)$$

This equation is derived from the experimentally observed [22] Star Formation Rate as a function of redshift [SFR(z)] as discussed in the appendix. In order to solve it we make the approximation that H(z) is given by the ΛCDM model, an assumption that is shown below to be accurate for low redshifts. The resulting $\Omega'_{reI}(z)$ function is plotted in Fig. 2 in blue (online version) or gray (in print). A quick inspection shows that it can be fitted by a super Gaussian function $\Omega_{reI}(z) = \Omega^0_{reI} \cdot \text{Exp}[-z/z_1]^m$ commonly used to model fiber

     

optic digital pulses.[25] Here the curve fitting produces the following values for the three parameters: $\Omega_{rel}^0 = 1.577 \times 10^{-5}$, $z_1 = 2.689$ and $m = 2.128$ and $\Omega_{rel}(z)$ is also plotted in Fig. 2 in black. Notice that the small value of $\Omega'_{rel,0}$ (see estimation in the appendix) results in a small $\Omega_{rel}^0$. In equation (17) below $\Omega_{rel}^0$ will be allowed to vary as a fitting parameter, while $z_1$ and $m$ will be kept at their values above as they already reproduce the dark energy properties quite well. This will determine the dark energy density needed in the present model, so that it can be compared to the $\Omega'_{rel}(z)$ estimated above.

The super Gaussian curve fitting above has important practical applications, as it can be combined with equation (8) to yield the pressure and equation of state of the 'perfect fluid' $\Omega'_{rel}(z)$:

$$p_{rel}(z) = -\left[1 + \frac{m}{3} \cdot \left(\frac{z}{z_1}\right)^m \cdot \left(\frac{1+z}{z}\right)\right] \cdot \rho_{rel}(z) \tag{14}$$

$$w_{rel}(z) = -\left[1 + \frac{m}{3} \cdot \left(\frac{z}{z_1}\right)^m \cdot \left(\frac{1+z}{z}\right)\right] \tag{15}$$

$\rho_{rel}(z) = \rho_c \cdot \Omega_{rel}(z)$. Again notice that $\Omega_{rel}(z)$ produces negative pressure with $w_{rel}(z) < -1/3$ for all values of z in its domain ($0 \leq z \leq z_{rel}$). Therefore if this were the dominant energy density, it would produce accelerated expansion. It is useful to evaluate $w_{rel}(z)$ for some specific redshifts: $w_{rel}(0) = -1$, $w_{rel}(1) = -1.17$, $w_{rel}(1.4) = -1.30$ and $w_{rel}(2) = -1.57$.

### III.  THE $\Omega_s(z)$ STORED ENERGY MODEL

Fig. 3 shows the equations of state for matter, radiation and the three energy densities $\Omega_L(z)$, $\Omega_I(z)$ and $\Omega_{rel}(z)$ and now we group them together into a single function:

$$\Omega_s(z) = \begin{cases} \Omega'_{reI}(z) + \Omega_L(z) + \Omega_I(z), & 0 \leq z \leq z_{reI} \\ \Omega_L(z) + \Omega_I(z), & z_{reI} \leq z \leq z_r \\ \Omega_I(z), & z > z_r \end{cases} \qquad (16)$$

With $\Omega_L(z)$, $\Omega_I(z)$ and $\Omega'_{reI}(z)$ defined by equations (7), (12) and (13) respectively.

Although GR does not require energy conservation on a global scale, it does not explicitly forbid either the introduction of $\Omega_s(z)$ as a gravitating energy density. On the contrary: the application of GR to cosmology calls for the introduction of more matter/energy density than the baryonic matter density only, to provide solutions to Friedmann's equations that are in agreement with observational results. This implies a modification of the GR equations analogous to the introduction of the cosmological

                                                                     

constant or dark energy term as shown in eq. (01). However, in this model $\Lambda$ is not a new constant of nature; $\Lambda = \Lambda(z) = 8\pi G.\Omega_s(z)$ and evolves with redshift. The conjecture proposed here is that the energy density $\Omega_s(z)$ takes part in gravitation and represents energy flow from the matter to the geometry side of the GR equations, i.e., in the opposite direction to that proposed initially by Zeldovich. In essence $\Omega_s(z)$ is a scalar field coupled to the electromagnetic field of light from which it receives energy transfer via the photon redshifting process. This approach has two advantages as it gives a physical origin for dark matter and dark energy and it creates a straightforward connection between these densities and inflation.

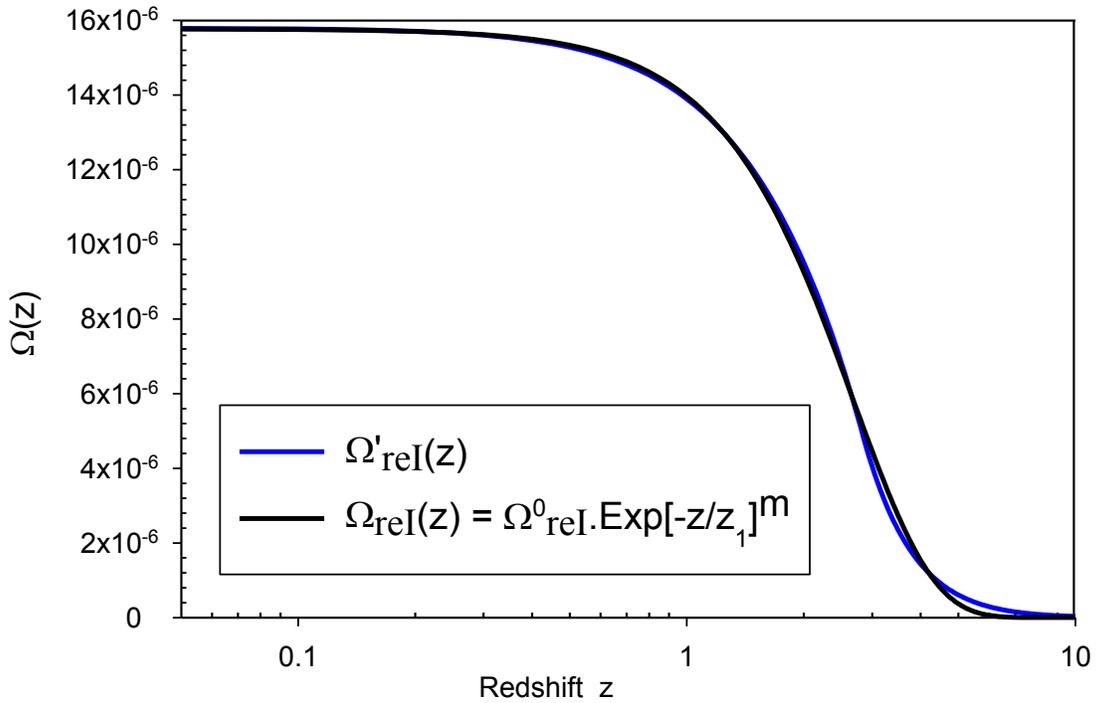

*Fig. 2 - (Color Online). The perfect fluid $\Omega'_{reI}(z)$ is plotted as a function of redshift. A super Gaussian function is fitted to $\Omega'_{reI}(z)$ and plotted in black.*

## IV.   LARGE SCALE STRUCTURE

It is now possible to construct a Hubble function H(z) for a flat (k = 0) FLRW universe that includes matter, radiation and the three components of the energy density $\Omega_s(z)$:

$$[E(z)]^2 = \left[\frac{H(z)}{H_0}\right]^2 = [\Omega_m^0 + \Omega_r^0.(1+z_r) + \Omega_I^0.(1+z_I)].(1+z)^3 + \Omega_{reI}^0.e^{-\left(\frac{z}{z_1}\right)^m} \quad (17)$$

In deriving the equation above we discarded a term in $\Omega_I^0.(1+z)^4$, as it is smaller than the $\Omega_I^0.(1+z_I).(1+z)^3$ term by a factor of $(1+z)/(1+z_I)$. For redshifts much smaller than $z_I$ this is a negligibly small number. In addition, the approximated super Gaussian function was used instead of the physically derived $\Omega'_{reI}(z)$. Instead of fixing the parameter $\Omega_{reI}^0$ to the estimated value mentioned before, here it is allowed to vary so that it provides the Friedmann's normalization: for z=0 we have E(0) = 1 and

$$\Omega_{rel}^0 = 1 - [\Omega_m^0 + \Omega_r^0 \cdot (1 + z_r) + \Omega_I^0 \cdot (1 + z_I)] \tag{18}$$

The term $\Omega_r^0 \cdot (1 + z_r) = 0.053$ depends only on parameters that have already been measured with good accuracy and was discussed previously. In eq. (17) the terms coming from $\Omega_L(z)$ and $\Omega_I(z)$ appear added to the baryonic matter term $\Omega_m^0$ having the same $(1+z)^3$ dependence, in close analogy with the $\Omega_{Barion}$ and $\Omega_{CDM}$ densities in the ΛCDM model. The implication is clear: only the sum inside the brackets in equation (18) is needed as a fitting parameter in this model and here one is also free to choose whatever combination of values for these parameters that is needed to satisfy all other cosmological observations.

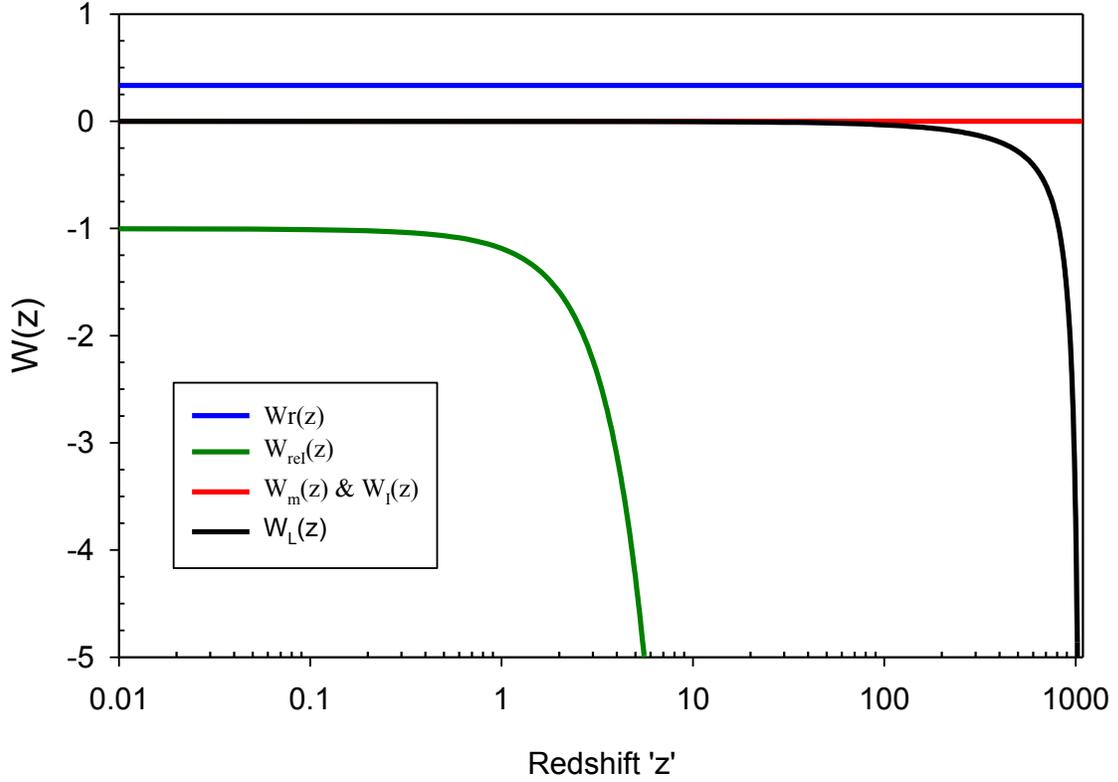

Fig. 3 - (Color Online). Equations of state for matter, radiation and the energy densities $\Omega_L(z)$, $\Omega_I(z)$ and $\Omega_{rel}(z)$.

### A. Fitting supernovae data

Assuming the FLRW metric and a flat universe as suggested by the WMAP7 results in Ref. [17], a Luminosity distance × redshift $D_L(z)$ function for standard candles can be constructed by canonical methods [10] and [26]:

$$D_L(z, \Omega_{mT}^0, \Omega_{reI}^0) = (1+z) \cdot \int_0^z \frac{dz'}{E(z', \Omega_{mT}^0, \Omega_{reI}^0)} \tag{19}$$

The function $E(z', \Omega_{mT}^0, \Omega_{reI}^0)$ is given by equation (17) and the two fitting parameters are the total 'matter' term $\Omega_{mT}^0 = [\Omega_m^0 + \Omega_r^0 \cdot (1+z_r) + \Omega_I^0 \cdot (1+z_I)]$ and $\Omega_{reI}^0$, the redshifted energy accumulated since re-ionization so that $\Omega_{mT}^0 + \Omega_{reI}^0 = 1$. Then the magnitude-redshift relation m(z), which is the measured quantity in SNIa tests, is given by:

$$m(z) = M + 5 \log[D_L(z, \Omega_{mT}^0, \Omega_{reI}^0)] \tag{20}$$

M = ($M_{abs}$ − 5 log$H_0$ − 25) is a third fitting parameter related to the absolute magnitude $M_{abs}$ and the Hubble constant $H_0$. Equations (19) and (20) allow the first test of the model presented here, by fitting them to the Union2 Supernovae data [21], with the three fitting parameters $\Omega_{mT}^0$, $\Omega_{reI}^0$ and $M_{abs}$. Fig. 4 shows the Union2 SNIa data points with error bars and the result of the curve fitting is shown in black. The values obtained for the fitting parameters are M = 25.057, $\Omega_{mT}^0 = 0.294$ and $\Omega_{reI}^0 = 0.706$. Also shown in red

                                                         15

(online) or gray (in print) for comparison is the best ΛCDM fit to the data, with $\Omega_m^0 = 0.28$ and $\Omega_\Lambda = 0.72$ and it is clear that the $\Omega_s(z)$ 'stored energy' model presented here produces a fit to the SNIa data as good as (overlapping) the one provided by the ΛCDM model.

Having obtained $\Omega_{mT}^0 = 0.294$ and assuming the baryonic matter value of $\Omega_m^0 = 0.046$ as in the ΛCDM model, it is now possible to calculate the energy density term coming from inflation. The result is $\Omega_I^0.(1 + z_I) = 0.195$. Therefore, all energy densities are now properly normalized at z = 0 and are plotted in Fig. 5.

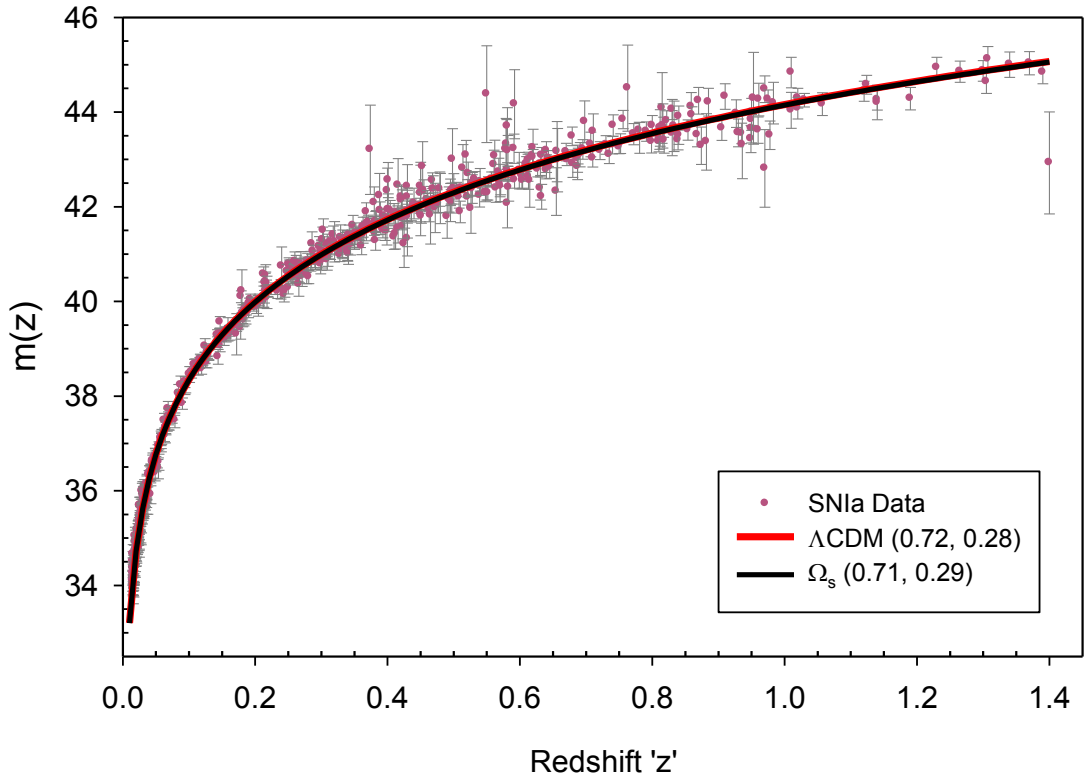

*Fig. 4 - (Color Online). Hubble diagram for the Union2 SNIa data with $\Omega_S$ curve fitting from eq.(20). The ΛCDM fitting is also shown for comparison. The numbers in parenthesis show the contributions of dark energy and matter terms respectively.*

                                                                                    16

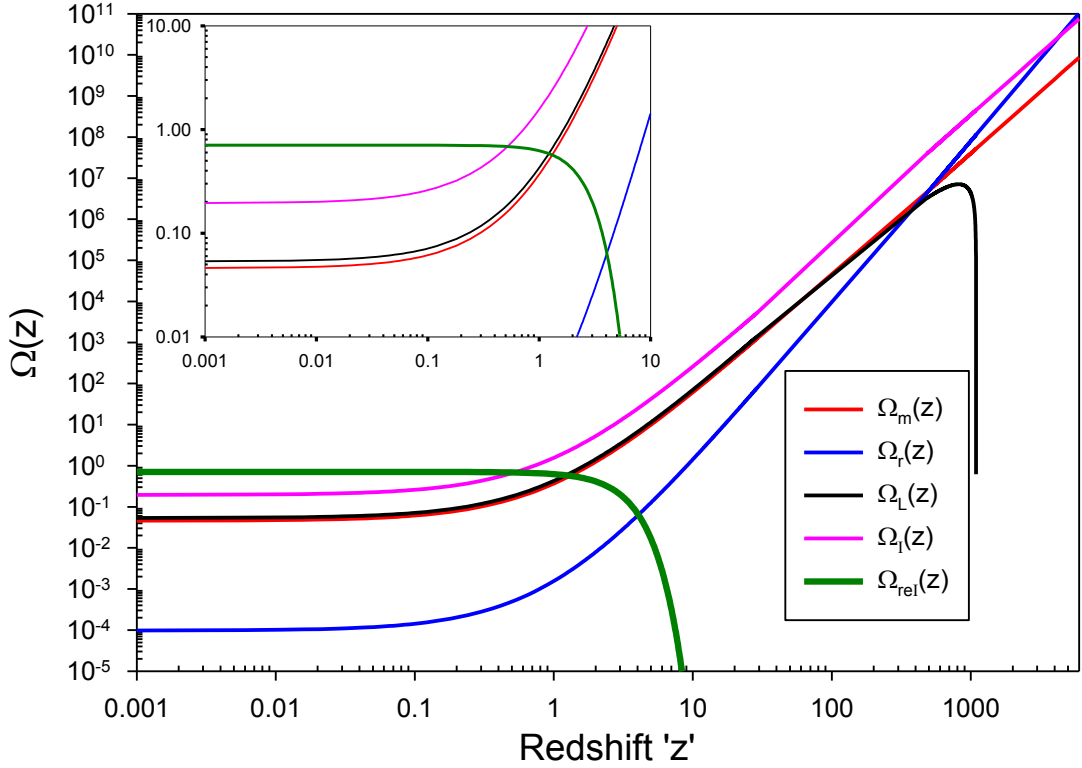

*Fig. 5 - (Color Online). The fractional energy densities in the universe as a function of redshift, normalized to the critical density at the present age. Notice the similar behavior of $\Omega_L$ and $\Omega_m$, for redshifts up to z ~ 300. The inset shows the same functions up to z = 10.*

### B. Look back time

Another important test for any cosmological theory is the calculation of the look back time, given by

$$t(z) = H_o^{-1} \cdot \int_0^z \frac{dz}{(1+z).E(z)} \qquad (21)$$

Where $H_0 = H(z=0) = (71\pm2.5)$ km.s$^{-1}$.MPc$^{-1}$ is the present time Hubble parameter. Taking the function E(z) with the values $\Omega_{mT}^0 = 0.294$ and $\Omega_{reI}^0 = 0.706$ from the SNIa fit and $H_0 = 71$ km.s$^{-1}$.MPc$^{-1}$ results in the t(z) function shown in Fig. 6. It gives the age of the universe as $t_U = 13.46\times10^9$ years, an age slightly lower than the one given by the

                                                                                         

ΛCDM model but perfectly compatible with the age of globular clusters, estimated to be larger than 11 Gyr.[27,28,29]  Other important ages and times are easily calculated as well. For example, in this model the recombination era happened at $t_{rec}$ = 399,433 years after the big bang, about ~20,000 years later than in ΛCDM.  In addition, assuming an 'effective' re-ionization redshift of $z_{reI}$ = 10, produces a re-ionization time of $t_{reI}$ = 464 million years after the big bang. In summary, the $\Omega_S(z)$ energy density in equation (16) fits the SNIa data as well as the ΛCDM model and produces ages that are compatible with the observational evidence from concordance cosmology. The age of the universe in particular has been a major stumbling block for other models that don't include some form of dark energy or cosmological constant.

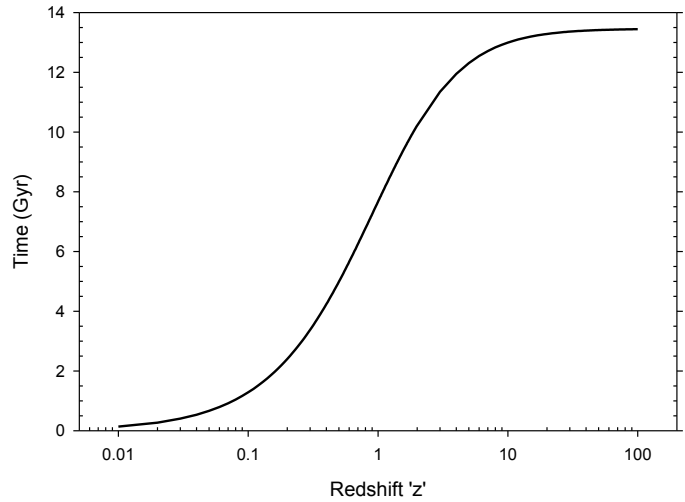

*Fig. 6 - Look-back time produced by equations (21) and (17), for a Hubble constant $H_0$ = 71 km.s$^{-1}$.MPc$^{-1}$. The age of the universe in this model is 13.46 Gyr and recombination happened ~399,433 years after the big bang.*

### C. Model predictions

                                                                                      

The $\Omega_S$ model might help explain some new puzzling difficulties with the ΛCDM model and offer some striking predictions. Recent observational evidence suggests that the ΛCDM model might have an 'age problem' after all [30], not for the present age, but for the age of the universe at high redshift. The old quasar "APM 08279+5255" has high metallicity and Fe/O = 3 ratio, which produces a best fit quasar age of at least 2.1 Gyr [31] after the big bang. On the other hand, based on the quasar redshift of z = 3.91, the ΛCDM age of the universe at the time of the quasar emission is only 1.7 Gyr. Several of the other cosmological models put forward to explain dark energy have been analyzed in this context as well and it was found that they cannot accommodate such an old quasar at this high redshift either.[30] At first sight the energy storage model presented here seems to have the same problem, as it produces an age of 1.56 Gyr at z = 3.91. However, in this case it is obvious where the problem lies: the isotropy and homogeneity assumptions embedded in the FLRW metric break down and the redshift to time conversion given by equations (17) and (21) is incorrect, because they take into account only the background energy terms. To these terms one has to add the contribution of the energy density transferred from the redshifted quasar photons and stored in the metric along the geodesic to us. This is still an FLRW type of metric, with an added contribution from the quasar similar to $\Omega'_{rel}(z)$ in the Hubble function (eq. 17). In essence the expansion along the geodesic illuminated by the quasar beacon is larger than the expansion produced by the background Hubble flow, resulting in a larger quasar redshift. Therefore, this model predicts that distances measured along a geodesic illuminated by a powerful beacon such as the old quasar "APM 08279+5255" might not be the shortest distances between two points. This extra expansion of the metric caused by the energy stored from the redshifted

  

photons is equivalent to a gravitational lensing effect, although the lens is *negative* in this case. In short, the quasar will be seen as having a larger redshift and therefore appear older than it really is, if only the background FLRW metric and resulting relation between time and redshift is used to calculate its age. Similar problems with the detection of an evolving dark energy when the homogeneity assumption breaks down have been noticed before and linear and non-linear lensing can lead to systematic errors in the distance-redshift relation that could distort results seriously.[32]

## V.    DISCUSSSION & SUMMARY

Having introduced the concept of an energy density $\Omega_s$ stored in the metric, we now review its properties and discuss its significance for dark energy/matter research. Starting with the component $\Omega_L(z)$, it produces negative pressure and equation of state with $w_L(0)$ = -0.0003, experimentally indistinguishable from non-relativistic matter. Acting on its own $\Omega_L(z)$ would produce accelerated expansion for redshifts in the range 544.5 < z < 1090, but for smaller redshifts the energy density dominates the pressure term, producing deceleration. If Friedmann's equation is solved with $\Omega_L(z)$ as the only density term, the time evolution of the scale factor is $a(t) \propto t^{2/3}$, which is a signature for the matter density. In addition $\Omega_L(z)$ enters the Hubble parameter only as a sum to the baryonic matter term. It does not have any thermal velocity (it is cold), does not emit any light (it is dark) and is diluted by the Hubble flow in the same way as matter.

The second component of $\Omega_S(z)$ is the newly defined $\Omega_I(z)$ term coming from inflation and it has the same properties as $\Omega_L(z)$. A large dark energy density at early times has been objectionable, as it would produce accelerated expansion and prevent element

                                                                                                       20

formation. In contrast, $\Omega_L(z)$ produces accelerated expansion only in the range $(z_I -1)/2 < z < z_I$, after which it behaves as matter, with the added attractive feature of providing a natural mechanism to turn OFF inflation. This might help explain why, despite over 7 decades of research, the mysterious dark matter particle has never been found. In its present form $\Omega_L(z)$ does not show how inflation is turned ON, which requires a dependence [33] of the scale factor on time at least as large as $a(t) \propto t^p$, with $p > 1$, whereas here we have $a(t) \propto t^{2/3}$. However, this need not be a serious difficulty; all that is needed is for photon production during inflation to be a dynamical process, with variations both in time and space. The spatial variations would result in $\Omega_I(z)$ spatial fluctuations that would have the same behavior as dark matter halos at late times. The conceivable time variations mean that the photons present during inflation did not come into existence all at once. A photon production process peaking sometime during inflation that mimics the shape of the UV luminosity of the universe since reionization would produce an accumulated energy density that is nearly constant for a short time and drive the exponential (or power-law) expansion of the scale factor.

$\Omega'_{rel}(z)$ is the last term of the stored energy $\Omega_S(z)$ and was derived as the energy transferred by photons produced and redshifted since the re-ionization era. It has a nearly constant density and an equation of state indistinguishable (within experimental uncertainties) from $w = -1$ for low redshifts where SNIa data are available. This energy density establishes a causal connection between the late time (low redshift) peak in the star formation rate and the concurrent emergence of dark energy. In the $\Lambda$CDM model this remarkable temporal overlapping is interpreted either as mere coincidence or by reference to some version of the anthropic principle.

                                                                                                                                                                                             21

We now show how the $\Omega_S$ model solves a problem with the $\Lambda$CDM cosmology that has not been discussed previously. The energy density demands made by the $\Lambda$CDM model seem to call into question its internal logical consistency. First it requires that the universe loses an unbounded amount of energy after the events of inflation, recombination and re-ionization through photon redshifting. As seen here, the term $\Omega_L(z)$ on its own is already larger than all the baryonic matter content of the universe today. After this demand for energy loss, the $\Lambda$CDM model then requires the introduction of two new energy densities as cold dark matter and dark energy that add up to 96% of the contents of the universe, in order to agree with observational results. Internal logical consistency seems to be undermined by these contradictory energy demands. In the $\Omega_S$ model this problem does not arise as the extra energy densities required to make it agree with observational results are in fact the same as those lost by the photon redshifting process.

The four main objections to the $\Lambda$CDM cosmology can be addressed by this model as well. Take first the *fine tuning problem*: the zero point energy of quantum fields is not needed here, as the $\Omega_S$ model is sufficient to explain the SNIa data. The natural value of the dark energy density is zero and the universe is relaxing to this asymptotic value as the expansion continues. The *coincidence problem* where the dark energy density coincides with the matter density within a factor of 2 or 3 is not of great significance here. The common origin of both dark matter and dark energy is the photon redshifting process in the $\Omega_S$ model. The third objection was that $\Lambda$CDM is in *direct conflict with the inflation.* $\Omega_S$ in contrast requires a powerful initial inflationary process with consequent photon redshifting to provide most of the energy density that is observed today as cold dark

                                                                                     22

matter. The fourth objection, that *no physical mechanism has been identified to explain dark energy* does not apply to this model. Here dark energy is seen as a scalar field coming from the photons produced and redshifted since re-ionization. Finally notice that the *why now* problem would be solved as well: a causal connection is established between the emergence of dark energy and the peak of star formation so that this is no longer viewed as a remarkable coincidence with an anthropic connotation.

New insights are needed to reconcile the small estimated value (see appendix) of $\Omega^0_{reI} = 1.577 \times 10^{-5}$, to the value needed to fit SNIa data, where $\Omega^0_{reI} = 0.706$. Our rough estimate of $\Omega^0_{reI}$ makes major simplifying assumptions that could be inaccurate by orders of magnitude, but at least it serves to identify the problem. The first light sources to ionize the universe, believed to be population III stars [34], have not been detected yet and there are complicating effects of feedback, the spectrum and fraction of the total number of photons that escape the forming galaxies [35,36], photon lifetime and light reprocessing in early galaxies, etc. These factors have been ignored in the estimate and if it proves to be reasonable, this would weaken the case for dark energy being the result of photon redshifting. Still, photon redshifting produces an energy density with the properties of dark energy and the amount produced ($1.6 \times 10^{-5}$) is much closer to what is required by $\Lambda$CDM (0.7) than the value given by vacuum energy ($>10^{120}$). A higher fidelity estimate would include stellar and galactic population synthesis and take into account the complicating factors above. Alternatively, improved understanding and technological advances might allow a direct measurement of the energy density removed from photons by redshifting since inflation, providing a direct test for this model.

 

Another issue with the $\Omega_S$ model was the use of $H_{\Lambda CDM}(z)$ to solve equation (13). A more rigorous approach would involve an iterative solution of equations (13) and (17) which gives H(z) for the $\Omega_S$ model. This might change slightly the values of the curve fitting parameters $\Omega_{mT}^0$ and $\Omega_{rel}^0$, but the changes are expected to be small because the two H(z) functions are identical at z = 0 and their difference increases to only 2.3% at z = 1.4, i.e., [$H_{\Lambda CDM}$ (1.4) - Hz(1.4)]/ $H_{\Lambda CDM}$ (1.4) = 0.023 and 9% at z=10, given the values of $\Omega_{mT}^0$ and $\Omega_{rel}^0$ obtained by fitting the SNIa data.

Ultimately a physical theory prevails if it is internally consistent, if it is consistent with the previous relevant body of knowledge and if it fits experimental/observational data. The present ΛCDM model fits the observations available at the expense of its internal consistency and leaves major problems unsolved. Here we provide a path to modify the ΛCDM model in a way that does not require the introduction of new forms of energy and that at the same time points to solutions to its four main objections.

Part of the writing and publication of this paper was supported by the JPL, Caltech, under a contract with NASA.

## VI. APPENDIX

A quick inspection of the rest frame Star Formation Rate (SFR) data in units of ($M_\odot.s^{-1}.MPc^{-3}$) in [22] reveals that it can be fitted by a power law:[36]

$$\text{SFR}(z) = \begin{cases} \beta \left[\frac{(1+z)}{(1+z^*)}\right]^\gamma, & z < z^* \\ \beta \left[\frac{(1+z)}{(1+z^*)}\right]^\delta, & z \geq z^* \end{cases} \quad (A1)$$

The curve fitting yields $\gamma = 2.544$, $\delta = -3.396$, $\beta = 5.868 \times 10^{-9}$ and $z^* = 2.837$. The number of solar masses per MPc$^3$ formed in the time interval (t, t+dt) is given by SFR(z).dt and a trivial conversion from time to redshift gives the number of solar masses formed per MPc$^3$ in the redshift interval (z, z+dz):

$$SFR(z).\frac{dz}{(1+z).H_0.E(z)} \tag{A2}$$

Where $H_0$ is in units of s$^{-1}$. During a star's lifetime a fraction $\alpha \cong 0.7\%$ of its core mass is converted to light by nuclear fusion.[37] Thus the function

$$f(z).dz = \alpha.M_\odot c^2.SFR(z).\frac{dz}{(1+z).H_0.E(z)} \tag{A3}$$

is an approximation for the light per MPc$^3$ produced by the star formation process from z to z+dz. We now make some assumptions to obtain the energy density lost to photon redshifting:

1 - Assume that all photons are emitted with the same energy $E_{em}$. A more accurate calculation would use stellar/galactic population synthesis and probability distribution functions for galactic spectra.

2 - We assume NO photon absorption, recycling or spectral evolution beyond photon redshifting caused by the expansion. In essence, all photons produced will contribute to the redshifting process.

Therefore $[f(z)/E_{em}].dz$ gives the number of photons emitted with energy $E_{em}$ in the redshift interval (z, z+dz), per MPc$^3$. Now, the energy lost by each photon between emission and detection at redshift "z" is:

　　　　　　　　　　　　　　　　　　25

$$\Delta E(z) = E_{em} - E_{det} = \frac{z}{(1+z)} E_{em} \quad (A4)$$

Finally the energy per MPc³ lost by photons emitted in the interval (z, z+dz) as they undergo redshifting is

$$\frac{f(z).dz}{E_{em}} \cdot \Delta E(z) = \alpha \cdot \frac{M_\odot.c^2}{H_0} \cdot SFR(z) \cdot \frac{z.dz}{(1+z)^2.E(z)} \quad (A5)$$

Integrating over the redshift range [z, z$_{reI}$] gives the energy density lost by the galactic photons through redshifting:

$$\Omega'_{reI}(z) = \alpha \cdot \frac{M_\odot.c^2}{H_0.\rho_c.c^2} \cdot \int_z^{z_{reI}} SFR(z) \cdot \frac{z}{(1+z)^2 E(z)} \, dz \quad (A6)$$

Where the energy density was normalized by the critical density $\rho_c.c^2$, with $\rho_c$ = 1.399×10$^{11}$ M$_\odot$.MPc$^{-3}$, $\alpha$ = 0.007 and H$_0$ = 2.3×10$^{-18}$ s$^{-1}$.

Combining equations (A1) and (A6) it is now possible to estimate the fractional density at present as

$$\Omega'_{reI}(0) = \Omega'_{reI,0} = 21754.7 \times \int_0^{z_{reI}} SFR(z) \cdot \frac{z}{(1+z)^2 E(z)} \, dz = 1.577 \times 10^{-5} \quad (A7)$$